\documentclass[8pt,a4paper]{article}
\usepackage[utf8]{inputenc}
\usepackage{amsmath,amssymb,amsthm}
\usepackage{graphicx}
\usepackage{hyperref}
\usepackage{geometry}
\usepackage{float}
\usepackage{titlesec}
\usepackage{authblk}
\usepackage{algorithm}
\usepackage{algpseudocode}

\titleformat{\section}[block]{\Large\bfseries}{\thesection.}{1em}{}
\usepackage[style=apa, backend=biber]{biblatex}
\DeclareLanguageMapping{english}{english-apa}
\newcommand{\chisq}[1]{\ensuremath{\chi^{2}_{#1}}}

\title{Maximum Likelihood Estimation and Likelihood Ratio Tests for the Power Spherical Distribution}
\author{Sharindi Amasha Samaraweera\textsuperscript{1} and Leif Ellingson\textsuperscript{1*}}
\affil{\small \textsuperscript{1}Department of Mathematics and Statistics, Texas Tech University, Lubbock, TX, USA}
\date{}

\begin{document}
\maketitle

\begin{abstract}

The Power Spherical distribution has been proposed as a computationally efficient alternative to the von Mises-Fisher distribution for data on the unit sphere, with reduced computational cost for simulation and a simpler normalizing constant. Despite these advantages, its statistical inference properties remain underdeveloped. In this paper, we develop a likelihood-based framework for estimation and inference for the Power Spherical distribution. Within this framework, we develop maximum likelihood estimation procedures for both the central direction and concentration parameter. Building on these estimators, we formulate likelihood ratio tests for hypotheses involving known and unknown parameters, study their asymptotic behavior, and evaluate finite-sample performance through simulation studies. The results demonstrate accurate Type I error control and good power when the asymptotic chi-square approximation is appropriate, while the parametric bootstrap provides reliable inference in small-sample settings or those where the dimension is large relative to the sample size. The findings also highlight nontrivial dependence of the power of tests for the central direction on the concentration parameter.
\\


\end{abstract}

\section{Introduction}

Probability distributions on unit spheres arise naturally in directional statistics \parencite{Mardia2009}, shape analysis via preshape representations of landmarks and curves \parencite{DrydenMardia2016}, and more broadly in the study of object data \parencite{PatrangenaruEllingson2016,MarronDryden2021}. A commonly used probability distribution for modeling data on the sphere that exhibit rotational symmetry about a central direction is the von Mises-Fisher distribution, often regarded as the spherical analogue of the normal distribution and studied extensively in both theoretical and applied contexts \parencite{Mardia2009, Banerjee2005}. Despite its appealing properties, the von Mises-Fisher distribution presents practical challenges, including numerical instability and computational cost arising from the modified Bessel functions in its normalizing constant \parencite{Sra2012}, as well as reliance on computationally intensive acceptance–rejection methods for simulation \parencite{Wood1994}. Motivated in part by these limitations in modern applications \parencite{Naesseth2017, Davidson2018}, \cite{DeCao2020} introduced the Power Spherical distribution as a computationally efficient alternative.

The Power Spherical distribution is isotropic about a central direction, but it offers notable computational advantages compared to the von Mises-Fisher distribution. In particular, it admits direct simulation and has a simpler normalizing constant, leading to substantial computational gains in large-scale settings. Motivated by these advantages, recent work has begun to incorporate the Power Spherical distribution into model formulation \parencite{Fortuin2022,Flaborea2024}. 

Despite this growing use, many of the key statistical characteristics of the Power Spherical distribution, including parameter estimation and formal inference, remain underexplored. In particular, likelihood-based estimation procedures and hypothesis testing methodologies for this distribution have not yet been systematically developed in the statistical literature, leaving important questions regarding estimation stability and inference unresolved. Likelihood-based inference provides a principled framework for statistical analysis in parametric models, enabling efficient parameter estimation and formal hypothesis testing \parencite{Casella2002}, and has played a central role in directional statistics through its application to classical models such as the von Mises-Fisher distribution \parencite{Mardia2009}. Extending these methods to the Power Spherical distribution is therefore essential for enabling rigorous statistical analysis beyond classical settings. 

In this paper, we address these gaps by developing a unified inferential framework for the Power Spherical distribution that includes stable maximum likelihood estimation procedures for both the central direction and concentration parameter as well as likelihood ratio tests for hypotheses involving known and unknown parameters. Estimation is carried out via a two-step iterative procedure, whose numerical behavior is examined across a range of parameter settings. For inference, asymptotic likelihood ratio tests based on chi-square approximations are developed, with a parametric bootstrap used to improve calibration in finite-sample or higher-dimensional settings. In addition, a data-driven rule is introduced for selecting the tuning constant used in the estimation of the central direction, improving stability and convergence. Simulation studies demonstrate the performance of the proposed methods in terms of estimation accuracy, Type I error control, and power, and highlight nontrivial dependence of the power of tests for the central direction on the concentration parameter.

The remainder of the paper is organized as follows. Section 2 introduces the Power Spherical distribution and reviews its defining properties, with emphasis on features relevant to likelihood-based inference. Section 3 develops maximum likelihood estimation procedures for the central direction and concentration parameters and discusses numerical aspects of stability and convergence. Section 4 presents likelihood ratio tests for hypotheses involving known and unknown parameters, including both asymptotic calibration and bootstrap-based alternatives. Section 5 reports simulation studies examining Type I error control, statistical power, and the effects of dimensionality and sample size. Section 6 concludes with a discussion of the main findings and directions for future research.

\section{The Power Spherical Distribution}

The Power Spherical distribution is defined on the unit sphere
\(\mathbb{S}^{d-1} = \{ x \in \mathbb{R}^d : \|x\|_2 = 1 \}\) and is parameterized by a central (modal) direction 
\(\mu \in \mathbb{S}^{d-1}\) and a concentration parameter \(\kappa \ge 0\). Let \(X\) be a random vector following the Power Spherical distribution with parameters \((\mu,\kappa)\). The probability density function  $p(x;\mu,\kappa)$ is given by
\begin{equation}\label{eq:ps_density}
p(x;\mu,\kappa)
=
\frac{1}{N(\kappa,d)}
\left(1+\mu^\top x\right)^{\kappa},
\text{for } x \in \mathbb{S}^{d-1}, \quad \text{where } 
N(\kappa,d) = 2^{d-1+\kappa}\pi^{\frac{d-1}{2}}
\frac{\Gamma\!\left(\frac{d-1}{2}+\kappa\right)}{\Gamma(d-1+\kappa)}.
\end{equation}
We note that the density depends on the observation $x$ only through the inner product $\mu^\top x$. The parameter $\kappa$ controls the concentration of the distribution about the central direction $\mu$: for large values of $\kappa$, the distribution is concentrated near $\mu$, while for smaller values it becomes more diffuse.

This behavior is reflected in the first moment of the distribution. The mean of the distribution in the ambient space $\mathbb{R}^d$ is given by
\begin{equation}\label{eq:ps_exp}
 \mathbb{E}[X] = \left( \frac{\kappa}{(d-1)+\kappa} \right) \mu.
\end{equation}
Thus, for $\kappa > 0$, the mean is a scalar multiple of $\mu$ that lies in the interior of the unit ball. As $\kappa \to 0$, $\mathbb{E}[X] \to 0$, and when $\kappa = 0$ the mean is the origin. In this case, $\mu$ is not uniquely defined, since the density depends on $x$ only through the inner product $\mu^\top x$. Further, as discussed by \cite{DeCao2020}, the distribution reduces to the uniform distribution on $\mathbb{S}^{d-1}$ when $\kappa = 0$. Accordingly, inference for $\mu$ is restricted to the case $\kappa > 0$.

\section{Maximum Likelihood Estimation}

We consider maximum likelihood estimation of the parameters $(\mu,\kappa)$ based on independent observations $x_1,\dots,x_n \in \mathbb{S}^{d-1}$ from the Power Spherical distribution. The likelihood function is given by
\begin{equation*}\label{eq:ps_Lik}
L(\mu,\kappa\mid x_1,\dots,x_n)
= N(\kappa,d)^{-n}\prod_{i=1}^n \bigl(1+\mu^\top x_i\bigr)^\kappa.
\end{equation*}

For numerical stability, we work with the log-likelihood
\begin{equation}\label{eq:ps_logLik}
\ell(\mu,\kappa)
= -n\log N(\kappa,d) + \kappa\sum_{i=1}^n \log\bigl(1+\mu^\top x_i\bigr).
\end{equation}

Maximization of the log-likelihood presents distinct challenges for $\mu$ and $\kappa$. The parameter $\mu$ is constrained to lie on the unit sphere, leading to a constrained optimization problem, while $\kappa$ enters the likelihood through both the normalizing constant and the log terms, resulting in a nonlinear estimation problem. In addition, when $\kappa = 0$, the distribution reduces to the uniform distribution on $\mathbb{S}^{d-1}$, so that $\mu$ is no longer part of the model. The estimation procedures developed in this section therefore focus on the setting where $\kappa > 0$, where both parameters are well-defined. Estimation is carried out using separate procedures for $\mu$ and $\kappa$, which are described in the following subsections.

\subsection{Estimation of the central direction}

For fixed $\kappa \ge 0$, the log-likelihood in \eqref{eq:ps_logLik} depends on $\mu$ only through the second term, since the normalizing constant $N(\kappa,d)$ does not involve $\mu$. Maximizing $\ell(\mu,\kappa)$ with respect to $\mu$ is therefore equivalent to
\begin{equation*}\label{eq:mu_est1}
\max_{\mu \in \mathbb{S}^{d-1}}
\;
\sum_{i=1}^n \log\!\bigl(1+\mu^\top x_i\bigr).
\end{equation*}
Introducing a Lagrange multiplier to enforce the unit-norm constraint, differentiation yields the first-order condition
\begin{equation}\label{eq:mu_est2}
\sum_{i=1}^n \frac{x_i}{1+\mu^\top x_i}
=
\eta \mu,
\end{equation}
for some scalar $\eta$. This condition implies that $\mu$ is proportional to a weighted average of the observations, where the weights depend on $\mu$ through $(1+\mu^\top x_i)^{-1}$.

Equation \eqref{eq:mu_est2} suggests an iterative scheme in which the right-hand side is evaluated at the current iterate and the result is re-normalized to lie on $\mathbb{S}^{d-1}$. In practice, we employ a scaled fixed-point iteration of this form. To initialize the estimate for $\mu$, we obtain from \eqref{eq:ps_exp} the expression $\mu = \mathbb{E}[X]/\|\mathbb{E}[X]\|$, which corresponds to the extrinsic mean for data on the sphere \parencite{PatrangenaruEllingson2016}. This motivates the use of the extrinsic sample mean, given by the normalized sample mean, as a natural plug-in estimator to initialize the iteration.

Although the first-order condition \eqref{eq:mu_est2} does not depend explicitly on $\kappa$, the behavior of the iterative scheme is strongly affected by the concentration of the distribution. When $\kappa$ is small, the distribution is diffuse and the central direction is only weakly determined, which leads to both statistical difficulty in estimating $\mu$ and numerical instability in the fixed-point updates. A step-size parameter $c>0$ is therefore included to improve stability of the iteration. Smaller values of $c$ yield more stable updates, while larger values can accelerate convergence but may lead to instability in low-concentration settings. The choice of $c$ can be selected as a function of the concentration parameter.

\begin{equation*}\label{eq:opt_const}
\text{c}_{\text{opt}}(\kappa) =
\begin{cases}
6.5118 \times 10^{-4}, & \text{if } 0.01 \leq \kappa \leq 1, \\
\sqrt{\frac{0.0096}{1 + \exp(-11.2896(\kappa - 1.8882))}}, & \text{if } 1 < \kappa \leq 2.5, \\
0.1, & \text{if } \kappa > 2.5.
\end{cases}
\end{equation*}
Further details on the concentration-driven selection of $c$ are provided in Appendix~A. The resulting procedure is summarized in Algorithm~\ref{alg:mu_estimation}. Convergence is typically rapid across a wide range of parameter settings.

\begin{algorithm}[H]
\caption{Scaled fixed-point estimation of the central direction $\mu$}
\label{alg:mu_estimation}
\begin{algorithmic}[1]
\Require Data $x_1,\dots,x_n\in\mathbb{S}^{d-1}$; step-size $c>0$; tolerance $\varepsilon>0$; max iterations $K_{\max}$
\State Initialize $\mu^{(0)}\in\mathbb{S}^{d-1}$ (e.g., normalized sample mean)
\For{$k=0,1,\dots,K_{\max}-1$}
    \State $v^{(k)} \gets \sum_{i=1}^n \dfrac{x_i}{1+(\mu^{(k)})^\top x_i}$
    \State $\tilde{\mu}^{(k+1)} \gets \mu^{(k)} + c\, v^{(k)}$
    \State $\mu^{(k+1)} \gets \dfrac{\tilde{\mu}^{(k+1)}}{\|\tilde{\mu}^{(k+1)}\|_2}$
    \If{$\|\mu^{(k+1)}-\mu^{(k)}\|_2 < \varepsilon$}
        \State \textbf{break}
    \EndIf
\EndFor
\State \Return $\widehat{\mu} \gets \mu^{(k+1)}$
\end{algorithmic}
\end{algorithm}

\subsection{Estimation of the concentration parameter}

For a fixed $\mu \in \mathbb{S}^{d-1}$, the concentration parameter $\kappa$ is obtained by maximizing the log-likelihood in \eqref{eq:ps_logLik} evaluated at that value of $\mu$. Differentiation with respect to $\kappa$ yields the score function
\begin{equation*}\label{eq:kappa_est2}
f(\kappa)
=
-n\frac{\partial}{\partial \kappa}\log N(\kappa,d)
+
\sum_{i=1}^n \log\!\bigl(1+\mu^\top x_i\bigr).
\end{equation*}

The maximum likelihood estimator $\widehat{\kappa}$ is the solution of $f(\kappa)=0$, obtained via Newton-Raphson iteration. The estimate for $\kappa$ is initialized using a plug-in method-of-moments estimator based on $\mu$, given by
\[
\kappa^{(0)}
=
(d-1)\Big/ \left(\frac{\overline{x}^\top \mu}{\overline{x}^\top \overline{x}} - 1\right),
\]
where $\overline{x} = \frac{1}{n}\sum_{i=1}^n x_i$.

To enforce $\kappa \ge 0$, negative updates are projected to zero. The resulting procedure is summarized in Algorithm~\ref{alg:kappa_estimation}. Estimation of $\kappa$ is carried out conditionally on $\mu$, and this conditional structure is used throughout the paper.

\begin{algorithm}[H]
\caption{Newton-Raphson estimation of $\kappa$ conditional on $\mu$}
\label{alg:kappa_estimation}
\begin{algorithmic}[1]
\Require Data $x_1,\dots,x_n\in\mathbb{S}^{d-1}$; fixed $\mu\in\mathbb{S}^{d-1}$; tolerance $\varepsilon>0$; max iterations $K_{\max}$
\State Compute $\overline{x} = \frac{1}{n}\sum_{i=1}^n x_i$
\State Initialize
$
\kappa^{(0)}
\gets
(d-1)\Big/ \left(\frac{\overline{x}^\top \mu}{\overline{x}^\top \overline{x}} - 1\right)
$
\For{$k=0,1,\dots,K_{\max}-1$}
    \State Compute $f(\kappa^{(k)})$ and $f'(\kappa^{(k)})$
    \State $\kappa^{(k+1)} \gets \kappa^{(k)} - \dfrac{f(\kappa^{(k)})}{f'(\kappa^{(k)})}$
    \State $\kappa^{(k+1)} \gets \max\{0,\kappa^{(k+1)}\}$
    \If{$|\kappa^{(k+1)}-\kappa^{(k)}| < \varepsilon$}
        \State \textbf{break}
    \EndIf
\EndFor
\State \Return $\widehat{\kappa} \gets \kappa^{(k+1)}$
\end{algorithmic}
\end{algorithm}

\section{Inference using Likelihood Ratio Tests}

This section develops likelihood ratio tests for inference under the Power Spherical distribution. We consider hypotheses of the form
\[
H_0 : \theta \in \Theta_0
\quad \text{versus} \quad
H_A : \theta \in \Theta,
\]
where $\theta = (\mu,\kappa)$, with $\mu \in \mathbb{S}^{d-1}$ and $\kappa > 0$, and $\Theta_0$ corresponds to the parameter values specified under the null hypothesis. We first present the hypotheses of interest together with their associated likelihood ratio statistics. We then describe the methods used to determine rejection regions for these tests, including both asymptotic approximations and parametric bootstrap procedures.

\subsection{Likelihood Ratio Statistics}

For each hypothesis $H_{0,j}$, $j=1,\dots,5$, the likelihood ratio test statistic, which is a monotone transformation of the likelihood ratio, is
\[
\lambda_j = -2\{\log L(\widehat{\theta}_0) - \log L(\widehat{\theta})\},
\]
 where $\widehat{\theta}_0$ and $\widehat{\theta}$ denote the constrained and unconstrained maximum likelihood estimators, respectively. Larger values of $\lambda$ indicate greater evidence against the null hypothesis. In each case, the statistic is obtained by substituting the appropriate estimators into \eqref{eq:ps_logLik} and simplifying. 

In the expressions below, $(\widehat{\mu}, \widehat{\kappa})$ denote the unconstrained estimators, while $(\widehat{\mu}_0, \widehat{\kappa}_0)$ denote the estimators under the null hypothesis.  When a parameter is fixed under $H_0$, the corresponding estimator is replaced by its fixed value. We consider the following hypotheses:

\begin{enumerate}
\item \textbf{Testing $H_{0,1}: \mu = \mu_0$ with known $\kappa$.}
\[
\lambda_1
= -2\kappa\sum_{i=1}^n
\log\!\left(
\frac{1+\mu_0^\top x_i}{1+\widehat{\mu}^\top x_i}
\right).
\]

\item \textbf{Testing $H_{0,2}: \mu = \mu_0$ with unknown $\kappa$.}
\[
\lambda_2
=
2n\log\!\left(\frac{N(\widehat{\kappa}_0,d)}{N(\widehat{\kappa},d)}\right)
-2\left[
\widehat{\kappa}_0\sum_{i=1}^n\log\!\bigl(1+\mu_0^\top x_i\bigr)
-
\widehat{\kappa}\sum_{i=1}^n\log\!\bigl(1+\widehat{\mu}^\top x_i\bigr)
\right].
\]

\item \textbf{Testing $H_{0,3}: \kappa = \kappa_0$ with known $\mu$.}
\[
\lambda_3
=
2n\log\!\left(\frac{N(\kappa_0,d)}{N(\widehat{\kappa},d)}\right)
-2(\kappa_0-\widehat{\kappa})
\sum_{i=1}^n\log\!\bigl(1+\mu^\top x_i\bigr).
\]

\item \textbf{Testing $H_{0,4}: \kappa = \kappa_0$ with unknown $\mu$.}
\[
\lambda_4
=
2n\log\!\left(\frac{N(\kappa_0,d)}{N(\widehat{\kappa},d)}\right)
-2\left[
\kappa_0\sum_{i=1}^n\log\!\bigl(1+\widehat{\mu}_0^\top x_i\bigr)
-
\widehat{\kappa}\sum_{i=1}^n\log\!\bigl(1+\widehat{\mu}^\top x_i\bigr)
\right].
\]

\item \textbf{Testing $H_{0,5}: \mu = \mu_0,\ \kappa = \kappa_0$.}
\[
\lambda_5
=
2n\log\!\left(\frac{N(\kappa_0,d)}{N(\widehat{\kappa},d)}\right)
-2\left[
\kappa_0\sum_{i=1}^n\log\!\bigl(1+\mu_0^\top x_i\bigr)
-
\widehat{\kappa}\sum_{i=1}^n\log\!\bigl(1+\widehat{\mu}^\top x_i\bigr)
\right].
\]
\end{enumerate}

\subsection{Determination of Critical Values}

Let $\alpha \in (0,1)$ denote the desired significance level. For each hypothesis $H_{0,j}$, $j=1,\dots,5$, the rejection region is of the form $\lambda_j > \lambda_{\alpha,j},$
where $\lambda_{\alpha,j}$ denotes the critical value at level $\alpha$.
Two approaches are used to determine $\lambda_{\alpha,j}$.

\medskip
\noindent\textbf{Asymptotic critical values.}
The statistic $\lambda_j$ has an asymptotic chi-square distribution under $H_{0,j}$, with degrees of freedom determined by the difference in dimension between the alternative and null parameter spaces \parencite{Lehmann2005, Casella2002}. Here, $\mu$ lies on the unit sphere $\mathbb{S}^{d-1}$ and therefore contributes $d-1$ free parameters, while $\kappa>0$ contributes one. Since $\kappa$ is assumed to be strictly positive, the parameter space does not involve boundary points, so the standard likelihood ratio theory applies directly.

The degrees of freedom for the hypotheses in Section~4.1 follow from the number of free parameters being tested and are summarized in Table~\ref{tab:df-table}. The critical value $\lambda_{\alpha,j}$ is taken as the $(1-\alpha)$ quantile of the corresponding chi-square distribution.

\begin{table}[h]
\centering
\caption{Degrees of freedom (df) for asymptotic likelihood ratio tests.}
\label{tab:df-table}
\vspace{0.2cm}
\begin{tabular}{|c|c|}
\hline
\textbf{Hypothesis} & \textit{\textbf{df}} \\ \hline
H$_{0,1}$ & $d - 1$ \\ \hline
H$_{0,2}$ & $d - 1$ \\ \hline
H$_{0,3}$ & $1$ \\ \hline
H$_{0,4}$ & $1$ \\ \hline
H$_{0,5}$ & $d$ \\ \hline
\end{tabular}
\end{table}

\medskip
\noindent\textbf{Parametric bootstrap critical values.}
When the asymptotic approximation is unreliable, $\lambda_{\alpha,j}$ can instead be obtained using a parametric bootstrap \parencite{DavisonHinkley1997}. That is, for each hypothesis $H_{0,j}$, parameters are estimated under the null using the methods of Section~3. Synthetic data sets are then generated from the fitted null model, and the statistic $\lambda_j$ is recomputed for each bootstrap sample using the same estimation procedure as for the observed data. The resulting values provide an empirical approximation to the null distribution of $\lambda_j$.

The critical value $\lambda_{\alpha,j}$ is taken as the empirical $(1-\alpha)$ quantile of this distribution.

\section{Simulation Studies}

This section presents a series of simulation studies designed to evaluate the finite-sample performance of the proposed likelihood ratio tests under various settings of the Power Spherical distribution. The simulations examine the empirical distribution and power of the likelihood ratio test statistics, with performance assessed across a range of dimensions, sample sizes, and concentration levels.

All simulation studies in this section follow a common framework. For a given dimension $d$, sample size $n$, and parameter configuration $(\mu,\kappa)$, we generate $m$ independent datasets from the Power Spherical distribution. For each dataset, the relevant maximum likelihood estimators are computed under the null and alternative hypotheses using the procedures in Section~3, and the corresponding likelihood ratio test statistic is evaluated. 

Tests are performed at significance level $\alpha=0.05$. Unless otherwise specified, critical values are obtained from the asymptotic chi-square reference distribution described in Section~4. In settings where this approximation exhibits noticeable finite-sample discrepancies, critical values are instead obtained using the parametric bootstrap procedure. When the asymptotic approximation is accurate, the bootstrap approximation to the null distribution is correspondingly close, and the resulting test behavior is similar. Accordingly, the bootstrap-based procedure is examined in settings where the asymptotic test exhibits non-negligible deviations from its nominal behavior. Empirical Type I error rates and powers are then estimated as the proportion of rejections across the $m$ Monte Carlo replications.

\subsection{Type I Error Behavior}

The finite-sample accuracy of the asymptotic chi-square approximation for the likelihood ratio test of $H_{0,5}$ is assessed through a large-scale Monte Carlo study based on empirical Type I error rates. For each configuration, $m=100{,}000$ independent datasets are generated from the Power Spherical model under fixed null parameters $(\mu_0,\kappa_0)$, with dimensions $d\in\{3,10,50,100\}$, sample sizes $n\in\{10,20,\ldots,500\}$, and concentration levels $\kappa_0\in\{2,5,15\}$ representing low, moderate, and high concentration scenarios. 

\begin{figure}[h]
     \centering
    \includegraphics[width=15cm]{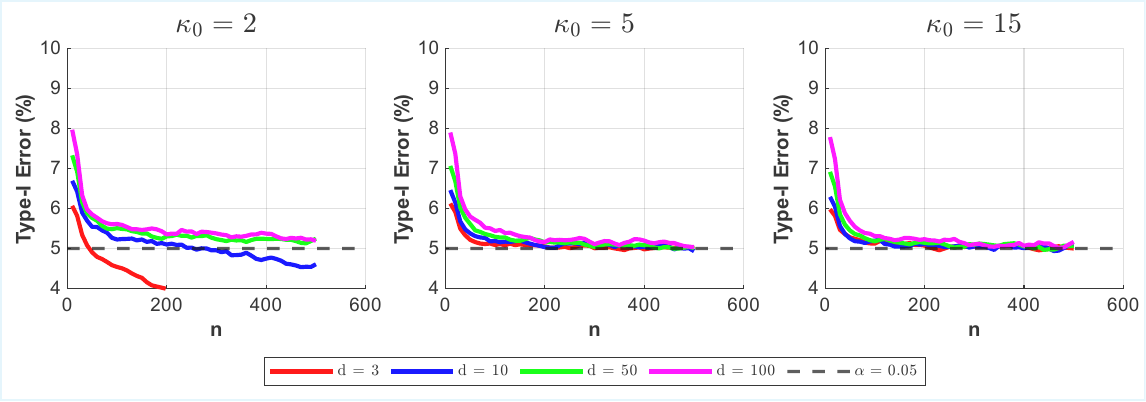}
    \caption{\centering Empirical Type I error rate for the likelihood ratio test for $H_{0,5}:\mu=\mu_0$ and $\kappa=\kappa_0$, evaluated across sample sizes and dimensions for $\kappa_0\in\{2,5,15\}$. Rejection is determined using critical values from the asymptotic $\chisq{d}$ reference distribution.}
    \label{fig:type1_joint_mu_kappa}
\end{figure}

Figure~\ref{fig:type1_joint_mu_kappa} reports the empirical Type I error rates for the asymptotic LRT across dimensions and sample sizes for $\kappa_0\in\{2,5,15\}$. The main deviation from nominal behavior occurs in the challenging setting consisting of low concentration and high dimension: when $\kappa_0=2$, rejection probabilities are inflated for small $n$, particularly for $d=100$ with $n\lesssim 100$. This reflects slower convergence of the LRT statistic to its chi-square limit in diffuse settings, where increased variability is consistent with weaker curvature of the likelihood function at low concentration. In contrast, for $d\in\{3,10\}$, Type I error stabilizes more quickly even under $\kappa_0=2$.

For $\kappa_0=5$, empirical rejection probabilities are close to nominal for moderate and large sample sizes across all dimensions, with only mild inflation when $n$ is small in the higher-dimensional settings. The most stable behavior is observed for $\kappa_0=15$, where Type I error approaches the nominal level rapidly across dimensions, consistent with the sharper likelihood geometry induced by high concentration. Overall, these results indicate that the asymptotic chi-square approximation provides reliable Type I error control when $n$ is sufficiently large relative to $d$, and that noticeable departures from nominal levels are primarily confined to higher-dimensional, low-concentration settings with small to moderate $n$.

\subsection{Power for Concentration Inference}

This section evaluates the empirical power of the asymptotic LRT for $H_{0,4}$. For each combination of true $\kappa \in \{0,...,50\}$, null $\kappa_0 \in\{5,15\}$,  $n \in \{20,50,100,200\}$, and $m=10{,}000$ Monte Carlo replications are generated, and power is estimated as the rejection proportion in the fixed dimension size $d=3$.

\begin{figure}[h]
    \centering
    \includegraphics[width=13cm]{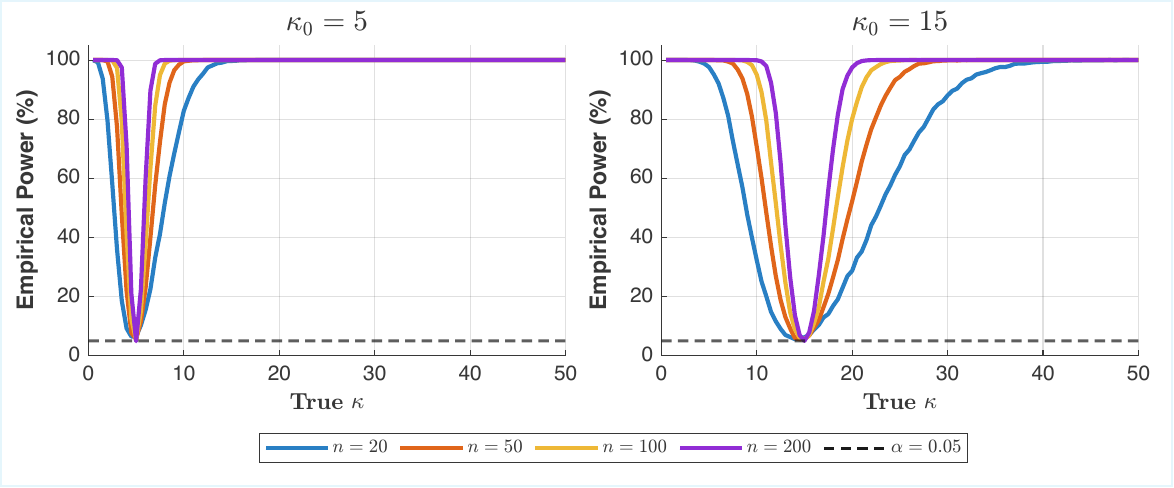}
    \caption{\centering Empirical power (\%) for the likelihood ratio test for $H_{0,4}:\kappa=\kappa_0$ when $\mu$ is unknown, shown for $d=3$ across sample sizes and $\kappa_0\in\{5,15\}$. Rejection is determined using critical values from the asymptotic $\chisq{1}$ reference distribution.}
    \label{fig:simulation2}
\end{figure}

Figure~\ref{fig:simulation2} shows that power increases as the true concentration deviates from $\kappa_0$, with steeper transitions as $n$ grows. When $\kappa_0=5$, the test exhibits a sharp increase in power away from the null, while maintaining a clear dip near $\kappa=\kappa_0$. For $\kappa_0=15$, the dip around the null is broader for small sample sizes, reflecting the greater difficulty of distinguishing nearby concentration values in an already highly concentrated scenario. Nevertheless, for $n\geq 100$, power approaches one across most of the range and remains close to the nominal level near $\kappa=\kappa_0$. These results indicate that the LRT is effective for detecting departures from $\kappa_0$ under unknown parameters, with performance improving as the sample size increases. However, the behavior of the test becomes more nuanced in higher-dimensional settings, which we examine next.

\subsubsection{Effect of Dimension on Power}
The impact of dimensionality on the power of the LRT is examined by extending the concentration test to higher-dimensional settings. Power curves are computed for moderate sample sizes $n\in\{20,50,100,200\}$ across dimensions $d\in\{10,25,50\}$ and null concentration values $\kappa_0\in\{5,15\}$.

\begin{figure}[h]
    \centering
    \includegraphics[width=12cm]{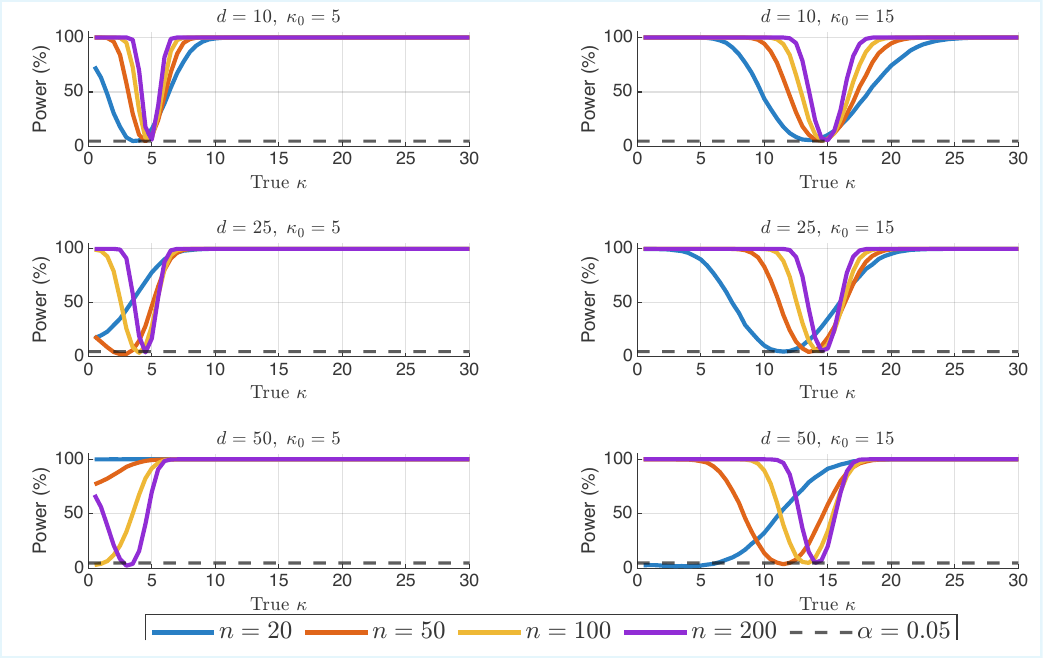}
    \caption{\centering Empirical power (\%) for the likelihood ratio test for $H_{0,4}:\kappa=\kappa_0$ when $\mu$ is unknown, shown in higher dimensions ($d\in\{10,25,50\}$) for moderate sample sizes ($n\in\{20,50,100,200\}$) and $\kappa_0\in\{5,15\}$. Rejection is determined using the asymptotic $\chisq{1}$ reference distribution.}
    \label{fig:power_moderate_n}
\end{figure}

As shown in Figure~\ref{fig:power_moderate_n}, the empirical power increases gradually as $\kappa$ departs from $\kappa_0$, but the rate of increase slows considerably as the dimension grows. In moderate dimensional settings (e.g., $d=10$), the rejection probabilities remain close to the nominal level near $\kappa=\kappa_0$ even at moderate sample sizes. In higher dimensions such as $d = 25$ and $d = 50$, the curves remain relatively flat across much of the $\kappa$ range, indicating reduced sensitivity of the test when the sample size is moderate. 

This behavior is particularly pronounced in low-concentration scenarios, where the distribution is more diffuse and the likelihood function is relatively flat in $\kappa$, making it more difficult to distinguish nearby concentration values. The effect is amplified as dimension increases, since the lack of concentration is reflected across more components, further reducing the discriminative ability of the test statistic. As a result, even when rejection probabilities near $\kappa=\kappa_0$ remain close to the nominal level, the power may increase only slowly as $\kappa$ departs from $\kappa_0$. For smaller values of $\kappa$, observations are more dispersed over the sphere, reducing the effective information available for distinguishing nearby concentration values. This effect becomes more pronounced as the dimension increases, since the signal associated with concentration is spread more thinly across the ambient coordinates. These patterns suggest that both distributional properties of the model and the dimensionality of the unit sphere influence the behavior of the test statistic, motivating a closer examination of the underlying chi-square approximation in these scenarios.

\subsubsection{Breakdown of the Asymptotic Approximation}

\begin{figure}[H]
    \centering
    \includegraphics[width=13cm]{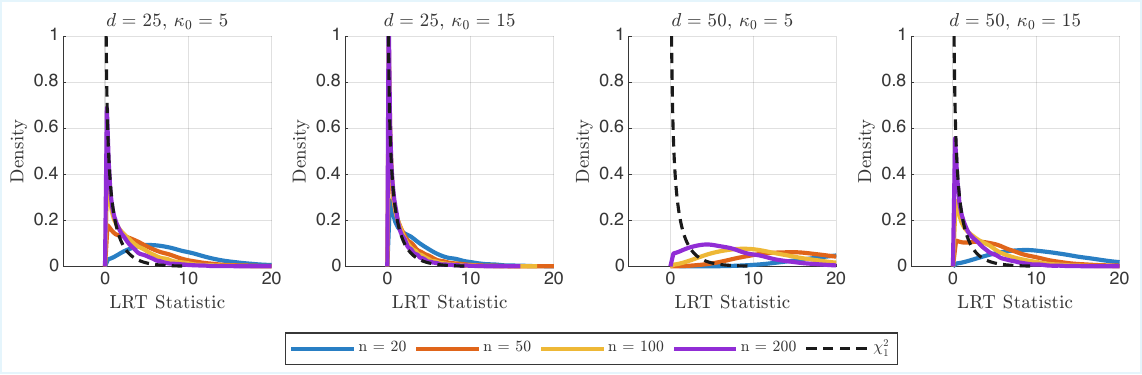}
    \includegraphics[width=13cm]{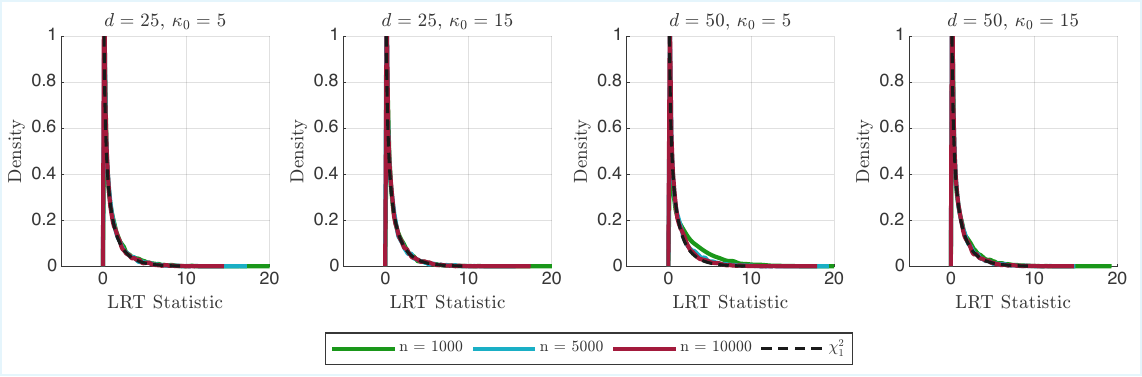}
    \caption{\centering Empirical null distributions of the likelihood ratio statistic for testing $H_{0,4}:\kappa=\kappa_0$ when $\mu$ is unknown, shown for $d\in\{25,50\}$ and $\kappa_0\in\{5,15\}$. The empirical densities are compared to the asymptotic reference density $\chisq{1}$.}
    \label{fig:high_n}
\end{figure}

To clarify whether the flattened power curves are primarily a consequence of reduced sensitivity or inaccuracies in the asymptotic approximation to the null distribution, the empirical null distribution of the LRT statistic is compared to its chi-square reference distribution under increasing dimensions. Results are shown in Figure~\ref{fig:high_n} for $d\in\{25,50\}$ under both $\kappa_0\in\{5,15\}$, with the top panels corresponding to moderate sample sizes ($n=20,50,100,200$) and the bottom panels corresponding to substantially larger sample sizes ($n=1000,5000,10000$).

For moderate sample sizes, Figure~\ref{fig:high_n} shows that the empirical null distribution departs increasingly from the chi-square reference as dimension grows. The discrepancy is most pronounced when $\kappa_0=5$, where the null distribution becomes more skewed and over-dispersed, indicating that the asymptotic chi-square approximation does not accurately describe the finite-sample behavior in these settings. In contrast, when the sample size is increased by orders of magnitude, the empirical distributions become substantially more concentrated and align closely with the chi-square density across both dimensions and both null concentration values. This comparison indicates that the loss of power observed in top panels is largely driven by inaccuracies in the asymptotic approximation to the null distribution rather than by an inherent lack of discriminative ability of the LRT.
Additional power results for substantially larger sample sizes are provided in Appendix~B, where the asymptotic behavior of the test is shown to stabilize even in the higher-dimensional settings.

\subsubsection{Bootstrap Inference in Higher Dimensions}

The preceding results motivate the use of the parametric bootstrap procedure. This approach is evaluated for $d\in\{25,50\}$ and $\kappa_0\in\{5,15\}$, first under the extreme scenario with  $n=20$ (where $n<d$), and then under moderate sample sizes $n\in\{50,100,150,200\}$. For each configuration, the bootstrap critical value is obtained from $B_{\text{null}}=10{,}000$ null replicates, and power is estimated using $m=10{,}000$ Monte Carlo replications under the alternative.

\begin{figure}[h]
    \centering
    \includegraphics[width=11cm]{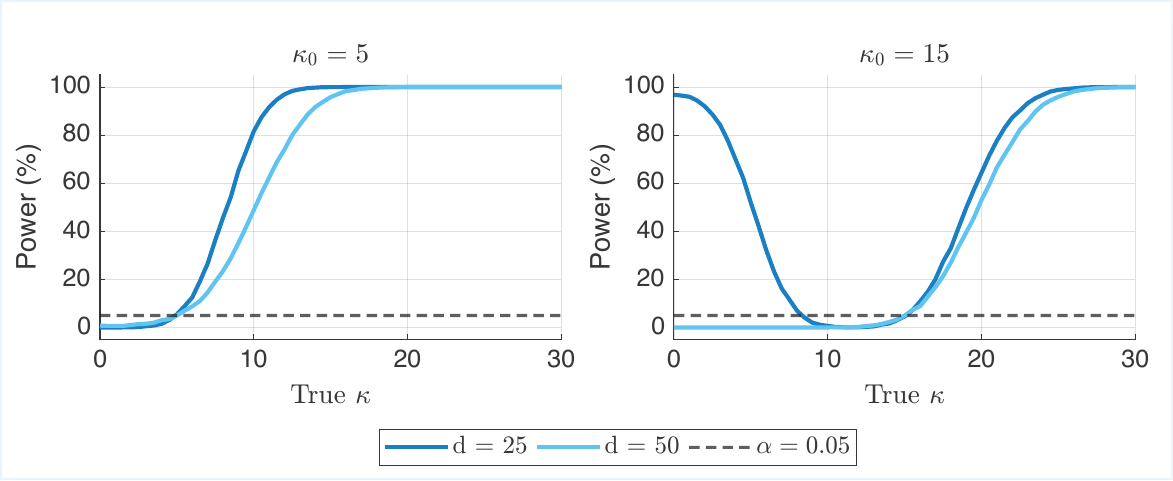}
    \caption{\centering Empirical power (\%) for the likelihood ratio test for $H_{0,4}:\kappa=\kappa_0$ when $\mu$ is unknown, using the parametric bootstrap critical value (based on $B_{\text{null}}=10{,}000$ null replicates). Results are shown for $n=20$ with $d\in\{25,50\}$ and $\kappa_0\in\{5,15\}$.}
    \label{fig:parametric_bootstrap_n_20}
\end{figure}

Figure~\ref{fig:parametric_bootstrap_n_20} shows that the bootstrap-based test produces well-shaped power curves even in the difficult scenario where $n<d$, recovering a clear dip near $\kappa=\kappa_0$ and a sharp increase away from the null. This indicates that the bootstrap-based procedure provides a more accurate finite-sample approximation in settings where the asymptotic chi-square approximation is less accurate. Mild flattening remains for the most challenging configuration ($d=50$, $\kappa_0=5$), suggesting that extremely diffuse distributions in very high dimensions may still require additional sample size for strong separation near the null.

\begin{figure}[h]
    \centering
    \includegraphics[width=11cm]{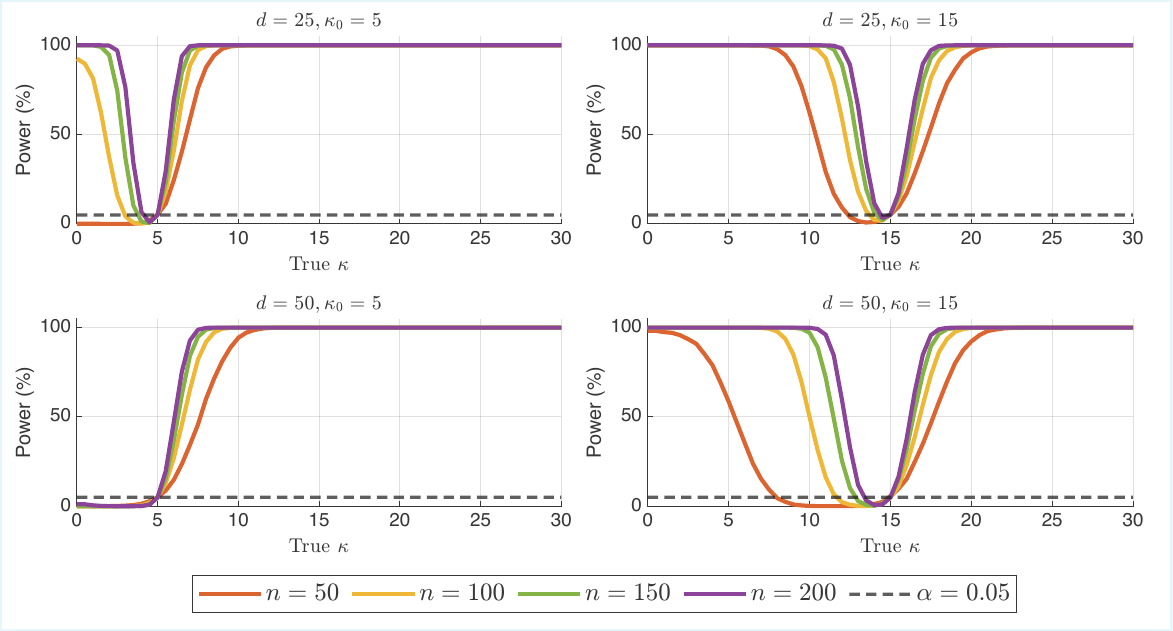}
    \caption{\centering Empirical power (\%) for the likelihood ratio test for $H_{0,4}:\kappa=\kappa_0$ when $\mu$ is unknown, using the parametric bootstrap critical value (based on $B_{\text{null}}=10{,}000$ null replicates). Results are shown for $d\in\{25,50\}$ across moderate sample sizes ($n\in\{50,100,150,200\}$) and $\kappa_0\in\{5,15\}$.}
    \label{fig:parametric_bootstrap_n_moderate}
\end{figure}

To further assess the effectiveness of the bootstrap in more practical sample size regimes, the analysis is repeated for moderate sample sizes $n \in \{ 50,100,150,200\}$ under the same combinations of $d$ and \(\kappa_0\). As shown in  Figure~\ref{fig:parametric_bootstrap_n_moderate}, the bootstrap yields consistently stable power curves once $n$ increases. In both $d=25$ and $d=50$, the transitions become steeper with larger $n$, Type I error remains close to the nominal level near $\kappa=\kappa_0$, and power approaches one as $\kappa$ moves away from the null. Overall, these results demonstrate that the parametric bootstrap provides an effective finite-sample correction in higher-dimensional settings, providing a more accurate finite-sample approximation to the null distribution and yielding interpretable power behavior when the chi-square approximation is inadequate.

Overall, these results show that the asymptotic LRT for concentration inference with unknown parameters performs well when $n$ is sufficiently large relative to $d$, but its power deteriorates in higher-dimensional settings with moderate or small sample sizes. In such scenarios, the empirical null distribution departs from its chi-square limit, producing flattened power curves and reduced sensitivity near $\kappa_0$, particularly under lower concentration. The parametric bootstrap provides a more accurate finite-sample approximation to the null distribution, yielding stable and interpretable power behavior when the asymptotic approximation is inadequate.

\subsection{Power for Central Direction Inference}
This section evaluates the empirical power of the asymptotic LRT for testing  $H_{0,2}$. Power is reported as a function of the directional shift $\|\mu-\mu_0\|$, which provides a direct geometric interpretation of the strength of the alternative. Simulations are conducted in dimension $d=3$. The true central direction is generated along a smooth path between the null direction $\mu_0=[0,1,0]$ and an orthogonal direction $M=[0,0,1]$,
\[
\mu(t)=\frac{(1-t)\mu_0+tM}{\|(1-t)\mu_0+tM\|},\qquad t\in[0,1],
\]
and power is computed over sample sizes $n\in\{50,100,200,1000\}$ and concentration levels $\kappa\in\{2,5,15\}$.

\begin{figure}[h]
    \centering
    \includegraphics[width=15cm]{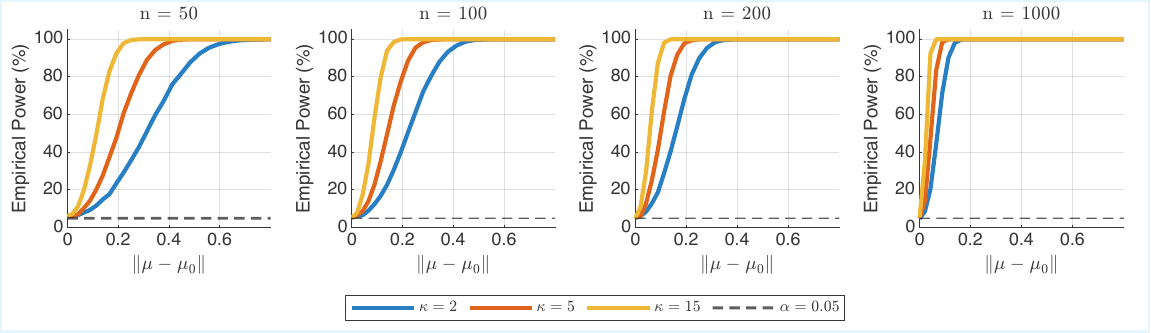}
   \caption{\centering Empirical power (\%) for the likelihood ratio test for $H_{0,2}:\mu=\mu_0$ when $\kappa$ is unknown, shown for fixed sample sizes ($n=50,100,200,1000$) across concentration levels $\kappa\in\{2,5,15\}$ in dimension $d=3$. Rejection is determined using critical values from the asymptotic $\chisq{d-1}$ reference distribution.}
    \label{fig:kappa_effect}
\end{figure}

Figure~\ref{fig:kappa_effect} illustrates the effect of concentration for fixed sample size. Increasing $\kappa$ produces substantially steeper power curves, indicating that more concentrated data yield stronger directional signal and allow smaller deviations from $\mu_0$ to be detected. In contrast, when $\kappa$ is small, the distribution is diffuse and larger directional shifts are required to achieve comparable power.
For each fixed sample size, the power curves become systematically steeper as $\kappa$ increases, and for larger $n$ the transition from low to high power occurs at smaller directional shifts.

\begin{figure}[h]
    \centering
    \includegraphics[width=15cm]{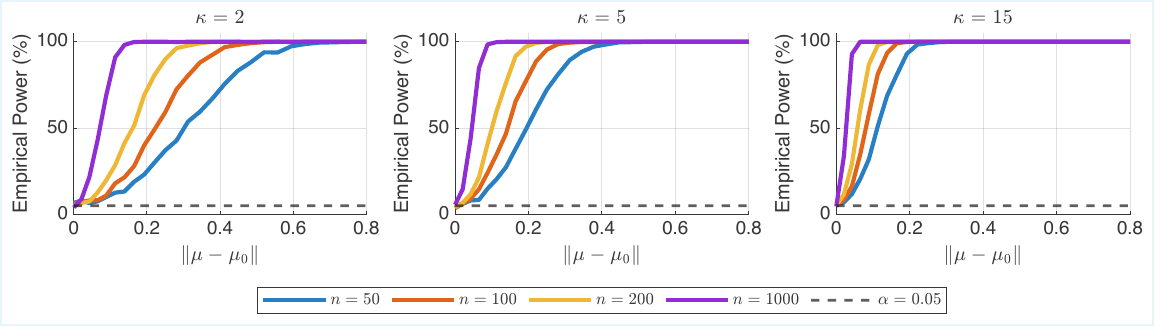}
    \caption{\centering Empirical power (\%) for the likelihood ratio test for $H_{0,2}:\mu=\mu_0$ when $\kappa$ is unknown, shown for fixed concentration levels ($\kappa=2,5,15$) across sample sizes ($n=50,100,200,1000$) in dimension $d=3$. Rejection is determined using critical values from the asymptotic $\chisq{d-1}$ reference distribution.}
    \label{fig:n_effect}
\end{figure}

Figure~\ref{fig:n_effect} shows the complementary effect of sample size for fixed concentration. For any $\kappa$, increasing $n$ yields uniformly higher power and a faster transition from rejection probabilities near the nominal level close to $\mu=\mu_0$ to near-perfect power as the shift increases. The improvement is most pronounced in the low concentration setting, where additional observations help compensate for the weaker directional structure.
For each fixed $\kappa$, increasing $n$ shifts the power curves upward and leftward, indicating that smaller directional deviations from $\mu_0$ can be detected as the sample size grows.

Overall, the LRT exhibits smooth and monotonic power behavior across all configurations, with the strength of directional signal governed jointly by $\kappa$ and $n$. In particular, lower concentration levels lead to more diffuse distributions and reduced sensitivity to deviations from $\mu_0$, while higher concentration enhances detectability of directional differences. These results indicate that reliable inference on the central direction remains feasible even when $\kappa$ is unknown, and that expressing alternatives in terms of $\|\mu-\mu_0\|$ provides a clear geometric description of test sensitivity.

\section{Discussion and Future Work}
\vspace{-0.2cm}
This paper presents a likelihood-based inferential framework for the Power Spherical distribution that combines stable maximum likelihood estimation with likelihood ratio testing and extensive simulation studies. A key contribution of the work lies in the development of a numerically stable estimation procedure for the central direction, together with an adaptive rule for selecting the update constant. This adjustment plays an important role in ensuring convergence across a wide range of dimensions, sample sizes, and concentration levels. When combined with Newton-Raphson estimation for the concentration parameter, the proposed approach leads to an estimation procedure that is both reliable and practical to implement.

The likelihood ratio testing framework introduced here allows for formal inference on both individual parameters and joint hypotheses within a unified setting. The simulation results show that, in situations where the chi-square approximation is appropriate, the proposed tests maintain good Type I error control and exhibit strong power. This behavior is most evident in low to moderate dimensions or when the sample size is sufficiently large relative to the dimension. In such settings, the Power Spherical model supports inference that is accurate and computationally efficient, further supporting its use as an alternative to the von Mises-Fisher distribution for directional data.

At the same time, the simulation studies reveal clear limitations of relying solely on asymptotic reference distributions in more challenging scenarios. In settings where the dimension is large relative to the sample size, particularly under low concentration where directional signal is weak, the empirical null distribution of the likelihood ratio statistic can differ noticeably from its theoretical chi-square limit. This discrepancy leads to distorted rejection probabilities and reduced power near the null, especially when distinguishing between nearby concentration values in diffuse settings. While such finite-sample deviations from asymptotic reference distributions are well known, the present results show how they arise in likelihood-based inference for the Power Spherical model through the interaction of concentration and dimension. An important observation from this study is that this behavior is largely driven by inaccuracies in the asymptotic reference distribution rather than by a fundamental weakness of the likelihood ratio test itself. When a parametric bootstrap is used instead, the resulting tests more accurately reflect nominal Type I error rates and yield smoother and more interpretable power curves, even when the dimension exceeds the sample size.

Several natural directions for future research emerge from this work. One possibility is the development of alternative inferential procedures, such as Wald type or score based tests, followed by a careful comparison of their finite-sample performance with likelihood ratio tests under higher-dimensional settings. Another extension involves multi-sample inference on the unit sphere, including problems that compare mean directions or concentration parameters across groups. Such developments would broaden the range of applications for which the Power Spherical framework can be used.

From a broader perspective, there is also scope to extend the Power Spherical model itself. Mixture models, hierarchical formulations, or related spherical families that preserve computational efficiency while offering additional flexibility are all promising directions. In addition, given the increasing use of spherical distributions in machine learning, future work could explore how the inferential tools developed here might be integrated into manifold based latent variable models, such as spherical variational autoencoders or flow based generative models. Bringing statistically principled inference into these settings would strengthen the connection between classical statistical methodology and modern representation learning on curved spaces.

Overall, the results of this paper demonstrate that the Power Spherical distribution supports not only efficient computation but also principled statistical inference. By addressing both estimation stability and hypothesis testing, particularly in settings where dimension exceeds sample size, this work provides a foundation for further theoretical development and for practical applications involving directional data.

\vspace{-0.3cm}
\section*{Conflict of Interest}
\vspace{-.1cm}
On behalf of all authors, the corresponding author states that there is no conflict of interest.

\vspace{-0.4cm}
%\newpage
\printbibliography

%\newpage
\appendix

\section{Concentration-Driven Selection of the Stability Constant}
\label{app:adaptive_c}

This appendix documents the empirical procedure used to construct the concentration-driven selection rule $c_{\mathrm{opt}}$ reported in the main text. The constant $c$ is used in the fixed-point update for the mean direction parameter $\mu$ and controls the step size of the iteration. Although $c$ does not appear in the likelihood, it strongly affects numerical stability of the MLE algorithm and therefore impacts estimation of $\kappa$ through its dependence on $\hat{\mu}$.

\subsection{Empirical Stability Study}

A simulation study is conducted to identify stable choices of $c$ across a broad range of parameter configurations. For each combination of $(\kappa,\mu,n)$, the MLE algorithm is evaluated over a grid of candidate values $c\in[0.001,0.1]$. The \textbf{maximum stable constant} is defined as the largest value of $c$ for which the iterative update for $\mu$ converges reliably. To obtain a conservative default choice that remains stable across settings, the minimum of these maximum stable values is taken over all configurations.

\setcounter{figure}{0}
\renewcommand{\thefigure}{A.\arabic{figure}}

\begin{figure}[h]
    \centering
    \includegraphics[width=11cm]{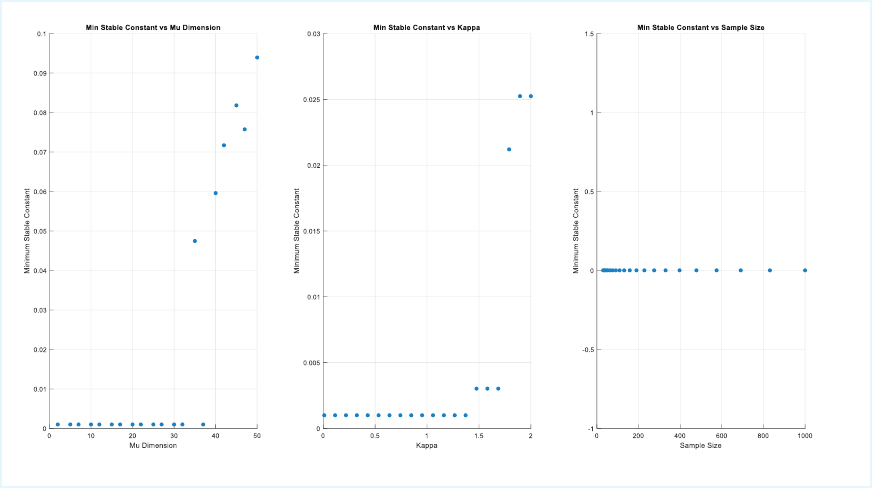}
    \caption{\centering Minimum of the maximum stable constant value across parameter configurations, shown as a function of dimension $d$, concentration $\kappa$, and sample size $n$. The stable constant is defined as the largest value for which the iterative MLE algorithm for $\mu$ converges reliably.}
    \label{fig:scatter1}
\end{figure}

Figure~\ref{fig:scatter1} summarizes the resulting stability boundary across the parameters $(d,\kappa,n)$. The dominant dependence is driven by $\kappa$, while the effects of dimension and sample size are comparatively smaller.

\begin{figure}[h]
    \centering
    \includegraphics[width=11cm]{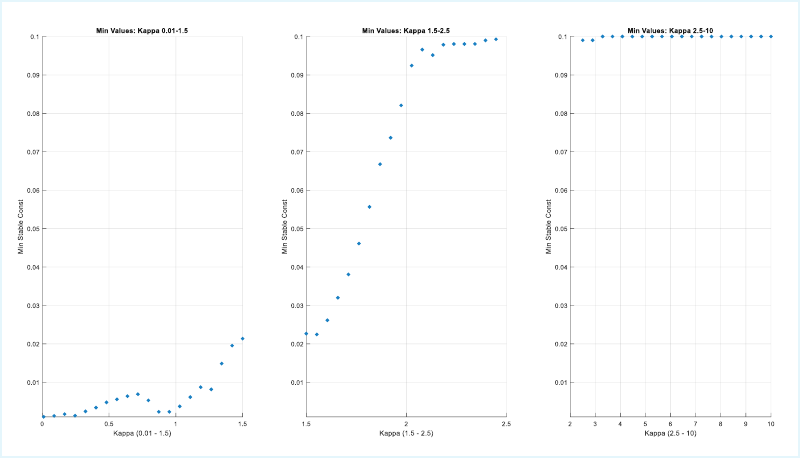}
    \caption{\centering Minimum of the maximum stable constant value as a function of $\kappa$, stratified into three scenarios: (left) $0.01 \le \kappa \le 1.5$, (middle) $1.5 < \kappa \le 2.5$, and (right) $\kappa > 2.5$. Each point corresponds to one simulated configuration, with $m=100$ replicates per setting.}
    \label{fig:scatter2}
\end{figure}
To examine the dependence on $\kappa$ more closely, the results are stratified into three concentration scenarios. Figure~\ref{fig:scatter2} shows clear phase-wise behavior: stable constants remain near a small baseline for low concentration, increase nonlinearly over an intermediate range, and approach the upper threshold for large $\kappa$. This pattern motivates a piecewise rule for selecting $c$ as a function of $\kappa$.

\begin{figure}[h]
\centering
\begin{minipage}{0.49\textwidth}
    \centering
    \includegraphics[width=\textwidth]{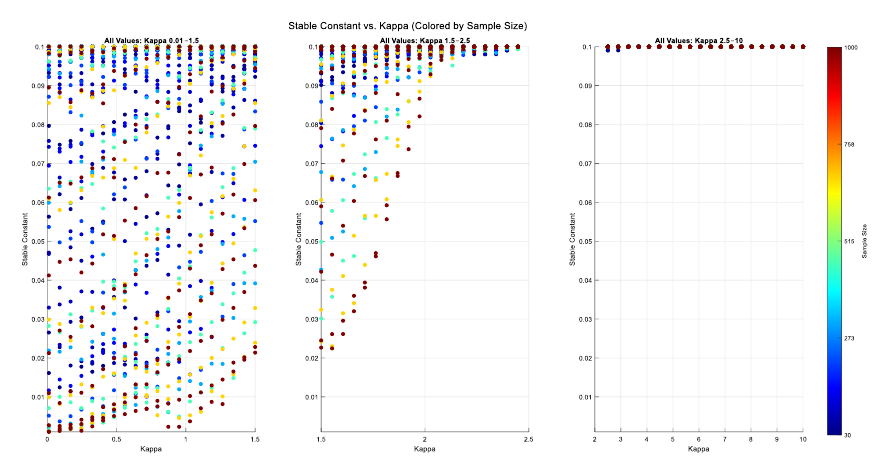}
    \caption{\centering Stable constant values versus $\kappa$, with color indicating sample size $n$, shown across the three $\kappa$ settings.}
    \label{fig:scatter3}
\end{minipage}
\hfill
\begin{minipage}{0.49\textwidth}
    \centering
    \includegraphics[width=\textwidth]{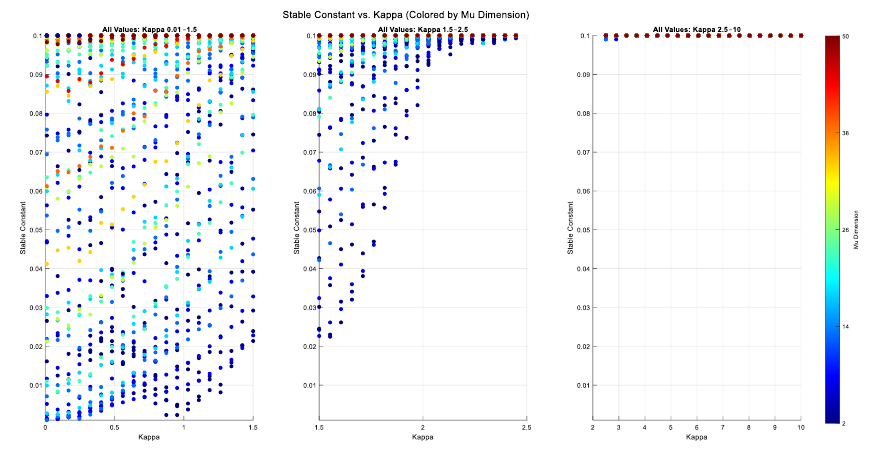}
    \caption{\centering Stable constant values versus $\kappa$, with color indicating the dimension $d$ of the unit sphere.}
    \label{fig:scatter4}
\end{minipage}
\end{figure}

Figures~\ref{fig:scatter3} and \ref{fig:scatter4} provide supporting evidence that the stability boundary is primarily governed by $\kappa$. Across all three scenarios, variation due to sample size and dimension is present but secondary relative to the overall $\kappa$ trend.

\subsection{Construction of the Concentration-Driven Rule $c_{\mathrm{opt}}(\kappa)$}

The stratified results in Figure~\ref{fig:scatter2} suggest three regimes. In the low concentration regime, stable constants remain close to a fixed baseline. In the high concentration regime, stable constants saturate near the upper threshold. The intermediate regime where $1<\kappa\le2.5$ exhibits nonlinear growth and is modeled using a logistic curve.

\begin{figure}[h]
    \centering
    \includegraphics[width=10cm]{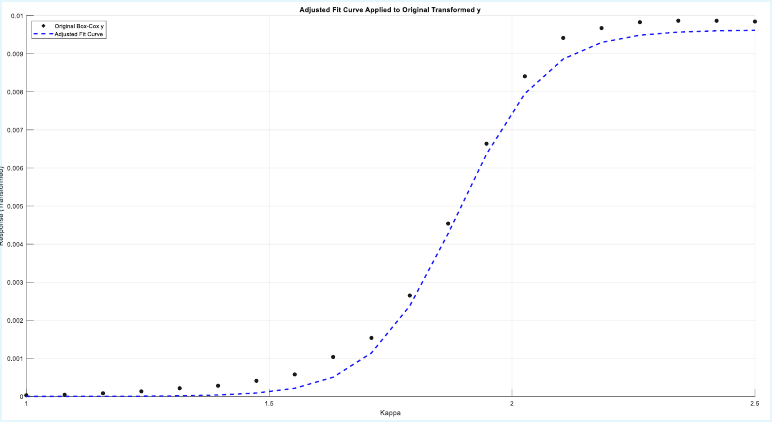}
    \caption{\centering Fitted curve for the intermediate regime $1<\kappa\le2.5$, shown on the Box-Cox transformed scale. The resulting model provides a smooth selection rule used to define $c_{\mathrm{opt}}(\kappa)$.}
    \label{fig:logistic}
\end{figure}

To improve fit in the intermediate regime, a Box-Cox transformation is applied to the response, and a conservative adjustment is introduced to reduce the risk of overestimating stable constants. The resulting fitted curve for the intermediate regime is shown in Figure~\ref{fig:logistic}.

Combining the three regimes produces the selection rule $c_{\mathrm{opt}}(\kappa)$ stated in the main text (Section 3.1.1), which is reproduced here for completeness. 

\[
c_{\mathrm{opt}}(\kappa)=
\begin{cases}
6.5118 \times 10^{-4}, & 0.01 \leq \kappa \leq 1, \\[6pt]
\sqrt{\dfrac{0.0096}{1+\exp\{-11.2896(\kappa-1.8882)\}}}, & 1<\kappa \leq 2.5, \\[10pt]
0.1, & \kappa>2.5.
\end{cases}
\]
This rule provides a stable and computationally efficient default choice of $c$ across a wide range of parameter configurations. Note that this rule is expressed as a function of $\kappa$, and in practice is implemented using an estimate of the concentration parameter. It is intended as a stability-oriented default rather than a formally optimal choice.

\section{Additional Power Analysis for Large Sample Sizes}

This appendix provides supplementary simulation results illustrating the behavior of the likelihood ratio test for testing $H_{0,4}:\kappa=\kappa_0$ with unknown $\mu$ under substantially larger sample sizes. These results extend the analysis in Section 5.2.2 by considering sample sizes $n\in\{1000,5000,10000,15000\}$ across dimensions $d\in\{10,25,50\}$ and null concentration levels $\kappa_0\in\{5,15\}$.

\setcounter{figure}{0}
\renewcommand{\thefigure}{B.\arabic{figure}}

\begin{figure}[h]
    \centering
    \includegraphics[width=12cm]{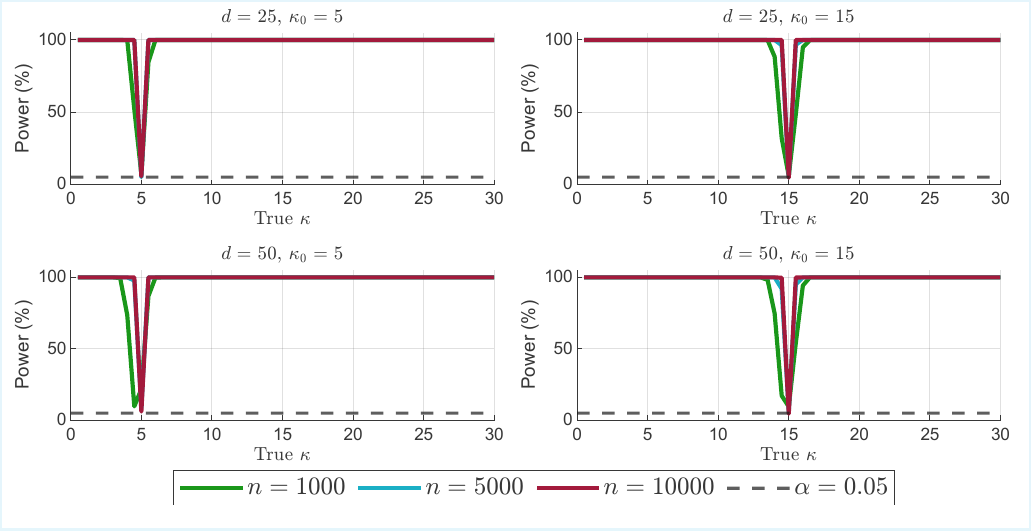}
    \caption{\centering Empirical power (\%) for the likelihood ratio test with null hypothesis $H_{0,4}:\kappa=\kappa_0$ with unknown $\mu$, shown for large sample sizes ($n=1000,5000,10000,15000$) across dimensions $d\in\{10,25,50\}$ and $\kappa_0\in\{5,15\}$. Rejection is determined using the asymptotic $\chisq{1}$ reference distribution.}
    \label{fig:power_large_n}
\end{figure}

Figure~\ref{fig:power_large_n} shows that increasing the sample size substantially improves empirical power across all dimensions, including higher-dimensional settings. Even for $d=50$, the power curves become steep and approach near-perfect power as $\kappa$ departs from $\kappa_0$. This behavior is consistent with the diagnostic results in Section 5.2.2, supporting the conclusion that the reduced power observed for moderate sample sizes in high dimensions is primarily driven by finite-sample discrepancies between the empirical null distribution and its asymptotic chi-square reference distribution, rather than by an inherent limitation of the likelihood ratio test.

\end{document}